\documentclass[12pt]{article}
\usepackage{times}
\usepackage{geometry}
\usepackage[utf8]{inputenc}
\usepackage{enumitem,amssymb}
\usepackage{natbib}
\usepackage{ragged2e}
\newlist{thematic}{itemize}{8}
\setlist[thematic]{label=$\square$}
\usepackage{pifont}
\usepackage{graphicx}        
\usepackage{color}           
\usepackage{breakurl}        
\usepackage{wrapfig}

\newcommand{\cmark}{\ding{51}}%
\newcommand{\done}{\rlap{$\square$}{\raisebox{2pt}{\large\hspace{1pt}\cmark}}%
\hspace{-2.5pt}}

\begin{document}
\raggedright
\huge
Astro2020 Science White Paper \linebreak
\LARGE
{\color{blue} Fundamental Physics with Radio Millisecond Pulsars}\linebreak

\normalsize
\vspace{-0.25cm}

\noindent \textbf{Thematic Areas:}\\ 
$\done$ Formation and Evolution of Compact Objects \hspace*{31pt} $\done$ Cosmology and Fundamental Physics \linebreak
  
\textbf{Principal Author:}  Emmanuel Fonseca (McGill Univ.), {\tt efonseca@physics.mcgill.ca} \linebreak
\textbf{Co-authors:} P.~Demorest (NRAO), S.~Ransom (NRAO), I.~Stairs (UBC), and NANOGrav

\vspace{0.5cm}
This is one of five core white papers from the NANOGrav Collaboration; the others are:
\begin{itemize}[noitemsep,topsep=0pt,leftmargin=*]
    \item \textit{Gravitational Waves, Extreme Astrophysics, and Fundamental Physics With Pulsar Timing Arrays}, J.~Cordes, M.~McLaughlin, et al.
	\item \textit{Supermassive Black-hole Demographics \& Environments With Pulsar Timing Arrays}, S.~R.~Taylor, S.~Burke-Spolaor, et al.
	\item \textit{Multi-messenger Astrophysics with Pulsar Timing Arrays}, L.Z.~Kelley, M.~Charisi, et al.
	\item \textit{Physics Beyond the Standard Model with Pulsar Timing Arrays}, X. Siemens, J. Hazboun, et al.
\end{itemize}


\vspace{0.5cm}

\justifying

\textbf{Abstract:}
We summarize the state of the art and future directions in using millisecond radio pulsars to test gravitation and measure intrinsic, fundamental parameters of the pulsar systems. As discussed below, such measurements continue to yield high-impact constraints on viable nuclear processes that govern the theoretically-elusive interior of neutron stars, and place the most stringent limits on the validity of general relativity in extreme environments. Ongoing and planned pulsar-timing measurements provide the greatest opportunities for measurements of compact-object masses and new general-relativistic variations in orbits.

\begin{center}
    \includegraphics[scale=0.1]{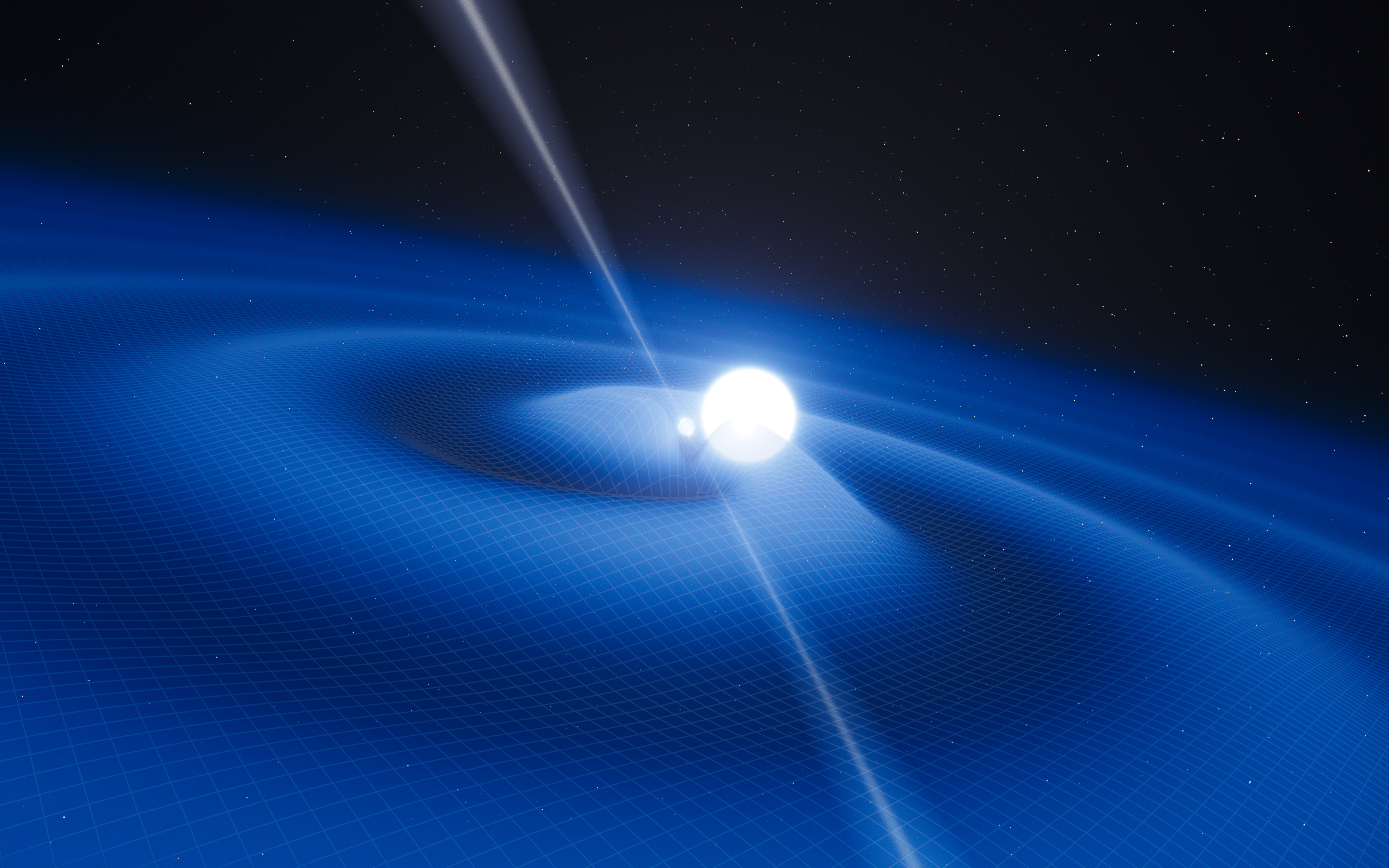} \\
    A radio pulsar in relativistic orbit with a white dwarf. (Credit: ESO / L. Cal\c{c}ada)
\end{center}

\pagebreak
\section{Key Motivations \& Opportunities}

One of the outstanding mysteries in astrophysics is the nature of gravitation and matter within extreme-density environments. Neutron stars especially pose a challenge as their central regions exceed the nuclear-saturation density, and therefore known nuclear physics, by factors of perhaps many.  Nuclear behavior at these densities allows for a wide range of compositions that ultimately produce different  ``equations of state" (EOSs), one of the fundamental relations that map  microphysical interactions to macroscopic quantities. Thus any constraints on the macroscopic parameters of neutron stars -- namely their masses and radii -- can strongly delimit the possibilities of internal structure and composition, and the strong gravitational forces holding them together.

Radio pulsars -- fast-spinning neutron stars emitting beamed radiation along their magnetic poles -- offer the greatest chances for high-precision measurements of relativistic effects and implied structural parameters. A classic example is the discovery and analysis of the ``Hulse-Taylor'' pulsar-binary system, where long-term timing provided the first solid (although indirect) evidence of quadrupole-order emission of gravitational radiation through the secular decay of its orbit \citep[e.g.][]{wh16}. Technological advances and ongoing pulsar discoveries are opening new avenues to strongly constrain the properties of ultra-dense matter and strong-field gravitation, especially in combination with future gravitational wave and x-ray observations. The requisite long-term, high-precision data sets can often serve dual purpose by contributing to gravitational wave detection using a pulsar timing array (see the white paper by Cordes and McLaughlin).

In this white paper, we highlight future opportunities and address the key questions that drive fundamental-physics studies of radio pulsars forward into the coming decade:

\begin{itemize}
    \item What is the EOS of bulk matter at the center of neutron stars?  And what are the maximum masses of those neutron stars?
    \item What is the correct theory of gravitation in the strong-field regime? At what point does Einstein's general relativity break down?
    \item With the dawn of gravitational-wave astronomy upon us, how can multi-messenger measurements of neutron stars be used to test gravitation and resolve the century-long mystery of their structure?
\end{itemize}

\section{Mass Measurements \& EOS Constraints} 
\label{sec:NS-mass}

The relativistic Shapiro time delay \citep{sha64} arises due to passage of pulsed emission through spacetime curvature induced by the companion star. In binary-pulsar timing models, the Shapiro delay is a function of two {\it a priori} unknown parameters, called the ``range" ($r$) and ``shape" ($s$) parameters. In most theories of gravitation, $s = \sin i$ where $i$ is the inclination of the binary system relative to the plane of the sky, whereas general relativity requires that $r$ = T$_\odot m_{\rm c}$, where T$_\odot = G{\rm M}_\odot/c^3$ and $m_{\rm c}$ is the companion mass.  By measuring these parameters, the pulsar mass ($m_{\rm p}$) is then immediately determined from the binary mass function. The Shapiro delay can be measured in a pulsar-binary system of any size, so long as the pulsar yields $\mu$s-level timing precision, and its inclination is favorable for significant measurement. 

Many of the 41 precisely-measured masses of neutron stars known to date\footnote{https://www3.mpifr-bonn.mpg.de/staff/pfreire/NS\_masses.html} have come from measurements of Shapiro delay alone \citep[see Table 2 of][]{of16}. Additional relativistic effects, such as periastron advance in eccentric binary systems, can be used to further improve the mass estimates \citep[see e.g.][]{fbw+11}. Models of the galactic millisecond-pulsar population predict the number of sub-$\mu$s-precision sources increasing by a factor of $\sim$5 over the next decade with next-generation radio telescopes (see the white paper by Lorimer et al.); the resulting mass measurements will number in the hundreds.  This development is paramount since new individual measurements, especially if extreme, still carry significant weight on a variety of compact-object studies.

\subsection{EOS Constraints from Pulsar Masses}

\begin{figure}
    \centering
    \includegraphics[scale=0.725]{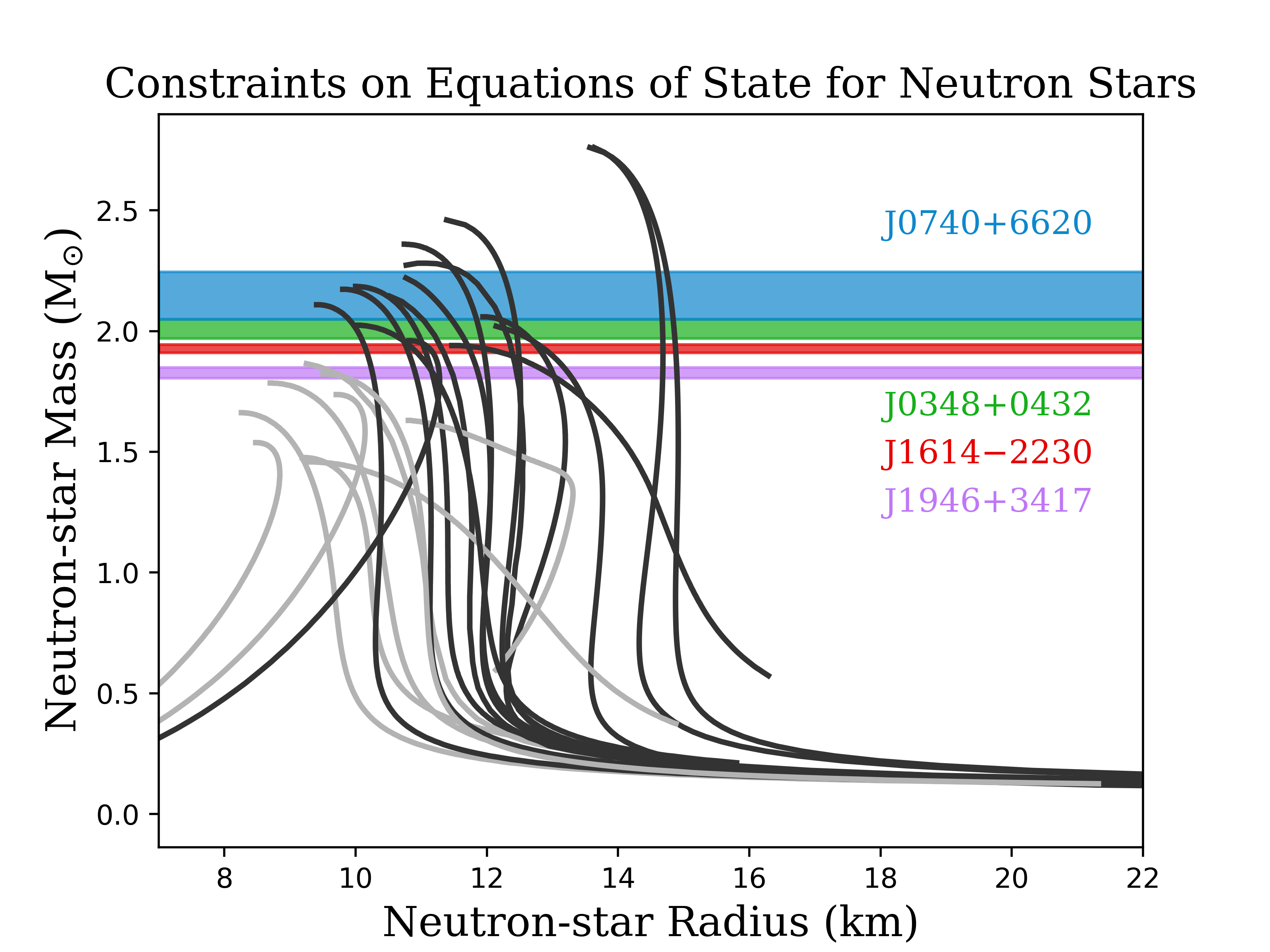}
    \caption{Mass-radius relations for different neutron-star equations of state shown by \citet{of16}; gray lines indicate EOS compositions that are excluded due to inconsistency with the J1614$-$2230 mass measurement. Horizontal lines indicate large-mass measurements made with high-precision pulsar timing.}
    \label{fig:my_label}
\end{figure}

Large pulsar masses test the viability of neutron-star equations of state, as each EOS differs in its prediction of maximum neutron-star masses. Figure 1 displays a collection of EOSs as tabulated by \citet{of16}, as well as four of the most massive pulsars currently known: J0348+0432 \citep{afw+13}; J0740+6620 (Cromartie et al., in prep); J1614$-$2230 \citep{fpe+16}; and J1946$+$3417 \citep{bfk+17}. The measurement of these pulsar masses, all around 2 M$_\odot$, eliminates nearly all equation-of-state models with strange quark and exotic hadron content.  Any increases in the maximum known neutron star mass -- even small increases -- will more stringently constrain equations of state.  Such increase may arise from future observations of the most massive known pulsars and, more importantly, future discoveries of new pulsars.

\subsection{Towards a New EOS Constraint: Moments of Inertia}
Classical variations of pulsar-binary parameters, such as orbital evolution due to spin-orbit coupling, require many decades to detect due to their smaller perturbations on pulse arrival times \citep{dt92}. However, variations from spin-orbit coupling yield unique information about the system, and especially on the pulsar's moment of inertia ($I_p$). 

Measurements of $I_p$ are unique since they, along with mass estimates, characterize internal properties of neutron stars and can be directly measured from pulsar timing \citep{ds88}. Due to the high degree of precision needed for this type of  measurement, long-term timing of known double-neutron-star systems yields the best opportunity for unambiguously constraining models of neutron-star interiors and their corresponding EOSs with measurements of $m_{\rm p}$ and $I_{\rm p}$. Based on current measurement precision and data span, two promising DNS systems for measurements of $I_{\rm p}$ are the double-pulsar system \citep{kw09} and PSR B1534+12 \citep{sttw02,fst14}.

\citet{rop16} demonstrated that a meaningful constraint on $I_p$ can be directly translated to constraints on the neutron-star radius, despite lingering uncertainties in assumed interior compositions that influence EOS forms. Thus a measurement of $I_p$ in a pulsar-binary system, coupled with the $m_{\rm p}$ estimates from relativistic-timing effects, will lead to a determination of preferred, finite regions in the mass-radius phase space shown in Figure 1.

\section{Tests of General Relativity} 
\label{sec:tests-of-GR}      

The traditional tests of gravitation with pulsars invoke self-consistency in predictions among the many ``post-Keplerian" effects observed in compact double-neutron-star systems \citep{dt92}. The current benchmark system for such tests is the relativistic ``double-pulsar" system, PSR J0737$-$3039A/B \citep{ksm+06}, which currently yields seven relativistic timing parameters and up to five tests of general relativity with agreement down to 0.05\% accuracy. While 15 double-neutron-star systems are currently known, these results are always dominated by the most extreme case that offers the best timing precision and conditions necessary for exquisite tests of general relativity.  Future discoveries of even tighter orbital systems will further test the validity of relativity theory to post-Keplerian orders greater than O$(v^2/c^2)$ in the ``strong-field" regime that is not accessible with Solar-System experiments.

\subsection{Testing the Universality of Free Fall}

New testing opportunities have arisen through a recent effort to understand the dynamics of PSR J0337+1715, a millisecond pulsar embedded within a compact three-body system with two low-mass white dwarfs \citep{rsa+14}. \citet{agh+18} were able to place a limit on a parameter that quantifies violation of the strong equivalence principle ($\Delta$), such that $|\Delta| < 2.6\times10^{-6}$ (with $\Delta = 0$ predicted by general relativity). This limit on $\Delta$ is the strongest yet obtained through experiment and will improve with observations taken over a long time span, especially using new ultra-wideband radio receivers currently in development.

Pulsar/white-dwarf binaries can also test for equivalence-principle violations, in the spirit of the parametrized post-Newtonian formalism of \citet{nw72}. These tests improve with the discovery of new binary pulsars, along with the formulation of additional preferred-frame tests \citep[e.g.][]{de92,swk18}

\subsection{Future Tests of Alternative Gravitational Theories}
A broad class of alternative theories invoke mediation of gravity through both tensor and scalar fields, whereas general relativity relies strictly on a tensor-field description. A key prediction from tensor-scalar theories is {\it dipolar} gravitational-wave radiation in compact binary systems with large differences in component binding energies \citep{ear75}. Pulsars in relativistic orbits with white dwarfs have already placed the strongest limits on tensor-scalar theories with constraints on dipolar radiation \citep[e.g.][]{fwe+12,zdw+19}. These constraints significantly improve with increasing pulsar timing baselines. 

Future improvements to tensor-scalar tests will also come in the discovery of pulsars tightly orbiting stellar-mass black holes. \citet{yl18} recently estimated that a small but detectable pulsar/black-hole binary population -- between 3 to 80 such systems -- resides within the Galactic disk. Ongoing and future pulsar-search programs are largely motivated by discovering these rare and extreme systems among the Galactic census of radio pulsars. Finally, pulsar-timing-array experiments are capable of detecting or constraining non-tensor polarization modes of gravitational waves (see the white paper by Siemens et al. for details).

\subsection{Future Tests with Galactic-Centre Pulsars}

A key science goal for future radio-astronomical observatories is the discovery and timing of radio pulsars in orbit around the supermassive black hole residing in the center of the Milky Way galaxy \citep[e.g.][]{bcc+18}. Recent projections have shown that an entirely new class of tests can be achieved with pulsar/black-hole orbits shorter than 1 yr in period, even if only one pulsar is discovered and yields low timing precision \citep{lwk+12}. These tests will directly probe mass and spin properties of the Galactic-centre black hole, as well as the validity of the storied ``no-hair" theorem \citep[e.g.][]{wil08} with measurements of mass, spin and the quadrupole term of its gravitational potential. 

\section{Synergy with Multi-wavelength/messenger Observations}

With the operation of new frontier observatories, a number of future opportunities will present themselves for unique constraints on neutron-star structure and tests of gravitation. In all these cases, radio timing measurements provide complementary information about these systems and are required to extract the best possible physical constraints:

\begin{enumerate}
    \item The Neutron Star Interior Composition Explorer ({\em NICER}) is currently measuring the thermal X-ray emission from millisecond pulsars with the goal to determine neutron star radii using relativistic effects.  For pulsars with known radio-determined masses, the constraints on the EOS will be much better.  Next-generation X-ray timing experiments with large collecting areas (e.g.~{\em STROBE-X} or {\em Athena}) will significantly improve such measurements. (See the whitepaper by Bogdanov et al. for related discussion.)
    
    \item Direct observations of gravitational waves from black-hole and neutron-star mergers with the Laser Interferometer Gravitational Wave Observatory (LIGO) are yielding new and independent tests of gravitation \citep{aaa+16}. Pulsar timing data and LIGO observations of neutron-star mergers \citep{aaa+18} provide complementary EOS constraints.
    
    \item Multi-wavelength observations of white-dwarf companions have had consequential impact on pulsar mass estimates through radial velocity measurements \citep[e.g.][]{afw+13}. The next generation of ground-based optical telescopes will significantly expand the volume of observable white-dwarf companions to millisecond pulsars.
\end{enumerate}

\section{Observational Requirements} 
\label{obs}

Dedicated timing measurements of radio pulsars in binary/triple systems continue to yield high-impact measurements of relativistic phenomena and structural neutron-star parameters. With fewer than 50 mass measurements made to date, the practice of probing fundamental physics with pulsars remains sensitive to individual discoveries and thereby an exciting prospect for the future; a single new pulsar-binary system can potentially alter our understanding of high-density nuclear processes, extreme-field gravitation, or limits on mass distributions and their relation to system-formation mechanisms. In order for these discoveries and measurements to continue, we need:

\begin{itemize}
    \item {\bf High-sensitivity, large-scale radio pulsar surveys}: New relativistic systems, including millisecond pulsar binaries, double neutron star systems, and triple systems, could provide much better pulsar masses and tests of gravity.  New and sensitive surveys of the radio sky are essential for such discoveries, and they have the potential to uncover tens of thousands of new pulsars, many of which will be unique and interesting (including pulsar/black-hole binaries).

    \item {\bf Targeted orbital-phase-specific observations}: \citet{dpr+10}, \citet{pen15} and Cromartie et al.~(in prep.) have shown that targeted observations at specific orbital phases can dramatically improve estimates of Shapiro-delay parameters (and thereby neutron star masses), especially when combined with a regular timing program.
 
    \item {\bf High-sensitivity and high-cadence observations}: Measuring the Shapiro delay requires high-precision timing and well-sampled orbits.  The first means using the best pulsars and the most sensitive GHz-frequency radio telescopes (imaging capabilities are not required), and the latter means regular and frequent timing observations over many orbits, and for complete timing solutions, over many years.(See the white paper by Lynch et. al.)
\end{itemize}

These requirements imply the need for large-area radio telescopes ($\gtrsim$100-m equivalent diameter) operating at $\sim$few hundred MHz--few GHz, able to dedicate substantial amounts of time (thousands of hours per year) to pulsar searching and timing observations over the next decade and beyond. Multiple instruments with these specifications would provide dramatic advances in the science described here, as well as in complementary goals such as pulsar-timing-array experiments.


\bibliographystyle{aasjournal}
\bibliography{journals,fundamental_physics_pulsars}  

\end{document}